\documentclass[12pt]{article}
\usepackage{amsmath,amssymb,epsfig,latexsym,txfonts,euscript}
\allowdisplaybreaks

\newcommand{\order}{{\cal O}}
\newcommand{\ps}{{p^2}}
\newcommand{\psp}{{p'^2}}

\newcommand{\sla}[1]{\rlap{\hspace{0.08cm}/}{#1}}
\def\nb{\bar{n}}
\def\X{{\EuScript X}}

\begin{document}

\begin{titlepage}

\begin{flushright}
CLNS~06/1954\\
FERMILAB-PUB-06-057-T\\[0.2cm]
March 17, 2006
\end{flushright}

\vspace{0.7cm}
\begin{center}
\Large\bf
\boldmath
Toward a NNLO calculation of the $\bar B\to X_s\gamma$ decay rate with a cut
on photon energy:\\
II. Two-loop result for the jet function
\unboldmath
\end{center}

\vspace{0.8cm}
\begin{center}
{\sc Thomas Becher$^a$ and Matthias Neubert$^b$}\\
\vspace{0.4cm}
{\sl $^a$\,Fermi National Accelerator Laboratory\\
P.O. Box 500, Batavia, IL 60510, U.S.A.\\[0.3cm]
$^b$\,Institute for High-Energy Phenomenology\\
Newman Laboratory for Elementary-Particle Physics, Cornell University\\
Ithaca, NY 14853, U.S.A.}
\end{center}

\vspace{1.0cm}
\begin{abstract}
\vspace{0.2cm}
\noindent 
The complete two-loop expression for the jet function $J(p^2,\mu)$ of 
soft-collinear effective theory is presented, including non-logarithmic terms. 
Combined with our previous calculation of the soft function $S(\omega,\mu)$, 
this result provides the basis for a calculation of the effect of a 
photon-energy cut in the measurement of the $\bar B\to X_s\gamma$ decay rate 
at next-to-next-to-leading order in renormalization-group improved 
perturbation theory. The jet function is also relevant to the resummation of 
Sudakov logarithms in other hard QCD processes.
\end{abstract}
\vfil

\end{titlepage}

\section{Introduction}

A significant effort is currently underway to complete the Standard Model 
calculation of the $\bar B\to X_s\gamma$ decay rate at next-to-next-to-leading 
order (NNLO) in renormalization-group improved perturbation theory. It is 
motivated by the fact that the relatively large branching ratio for this 
decay, combined with the increased precision in its measurement at the $B$ 
factories, make this an excellent way to probe for hints of new flavor 
physics. The experimental detection of $\bar B\to X_s\gamma$ events relies on 
the reconstruction of a high-energy photon, whose energy in the $B$-meson rest 
frame exceeds a value $E_0\approx 1.8$\,GeV. The theoretical analysis of the 
{\em partial\/} inclusive $\bar B\to X_s\gamma$ decay rate with a cut 
$E_\gamma\ge E_0$ must deal with short-distance contributions associated with 
three different mass scales: the hard scale $m_b$, an intermediate scale 
$\sqrt{m_b\Delta}$, and a soft scale $\Delta$, where 
$\Delta=m_b-2E_0\approx 1$\,GeV \cite{Neubert:2004dd}. The cut-dependent 
effects are described in terms of two perturbative objects called the jet 
function and the soft function, which for an analysis of the decay rate at 
NNLO are required with two-loop accuracy. The two-loop calculation of the soft 
function has been presented in \cite{Becher:2005pd}, while that of the jet 
function is described in the present work. As a by-product of our analysis we 
calculate the two-loop anomalous-dimension kernel of the jet function. 

The jet function $j(L,\mu)$ needed in the factorization formula for the 
partial $\bar B\to X_s\gamma$ decay rate \cite{Neubert:2005nt} is related to 
the original jet function $J(p^2,\mu)$ of a massless quark in QCD 
\cite{Sterman:1986aj} by 
\begin{equation}\label{jdefsdef}
   j\Big( \ln\frac{Q^2}{\mu^2},\mu \Big)
   \equiv \int_0^{Q^2}\!dp^2\,J(p^2,\mu) \,.
\end{equation}
While the perturbative expression for $J(p^2,\mu)$ involves singular 
distributions (see, e.g., \cite{Bauer:2003pi,Bosch:2004th}), the function $j$ 
has a double-logarithmic expansion of the form
\begin{eqnarray}\label{sexp}
   j(L,\mu) &=& 1 + \sum_{n=1}^\infty
    \left( \frac{\alpha_s(\mu)}{4\pi} \right)^n
    \left( b_0^{(n)} + b_1^{(n)} L + \dots + b_{2n-1}^{(n)} L^{2n-1}
    + b_{2n}^{(n)} L^{2n} \right) .
\end{eqnarray}
By solving the renormalization-group equation for the jet function order by 
order in perturbation theory, the coefficients $b_{k\ne 0}^{(n)}$ of the 
logarithmic terms in (\ref{sexp}) can be obtained from the expansion 
coefficients of the jet-function anomalous dimension and the $\beta$-function, 
together with the coefficients $b_0^{(n)}$ arising in lower orders 
\cite{Neubert:2005nt}. The two-loop calculation performed in the present work 
gives the constant $b_0^{(2)}$ and provides the first direct calculation of 
the two-loop anomalous dimension of the jet function. We also note that from 
our result for $j(L,\mu)$ one can derive the two-loop expression for 
$J(p^2,\mu)$ in terms of so-called star distributions 
\cite{Bosch:2004th,DeFazio:1999sv}.

We stress that even though our primary goal is to improve the theoretical 
analysis of $\bar B\to X_s\gamma$ decay, applications of our results are not 
confined to flavor physics. Indeed, the jet function is a universal object, 
which enters in many applications of perturbative QCD to jet physics, 
deep-inelastic scattering, and other hard processes. The two-loop calculation 
of the function $j(L,\mu)$ is described in Section~\ref{sec:2loop}. It follows 
closely our calculation of the soft function in \cite{Becher:2005pd}; however, 
the evaluation of the two-loop master integrals is considerably more 
complicated in the present case. In Section~\ref{sec:DIS} we briefly discuss 
jet-function moments and their renormalization-group evolution.

\section{Two-loop calculation of the jet function}
\label{sec:2loop}

The factorization properties of decay rates and cross sections for processes 
involving hard, soft, and collinear degrees of freedom become most transparent 
if an effective field theory is employed to disentangle the contributions 
associated with these different momentum regions. Soft-collinear effective 
theory (SCET) has been designed to accomplish this task 
\cite{Bauer:2000yr,Bauer:2001yt,Beneke:2002ph,Hill:2002vw}. In the context of 
SCET the jet function is defined in terms of the hard-collinear quark 
propagator \cite{Bosch:2004th,Bauer:2001yt}
\begin{equation}\label{eq:Jscet}
   \frac{\sla{n}}{2}\,\nb\cdot p\,{\cal J}(p^2,\mu)
   = \int d^4x\,e^{-ip\cdot x}\,\langle 0\,|\,{\rm T}\left\{
   \X_{hc}(0)\,\overline\X_{hc}(x) \right\}|\,0\rangle \,,
\end{equation} 
where $\mu$ is the renormalization scale, and $n$ and $\bar n$ are two 
light-like vectors satisfying $n\cdot\bar n=2$. For simplicity we suppress 
color indices on the quark fields. The propagator is proportional to a unit 
matrix in color space. The composite field 
$\X_{hc}(x)=S_s^\dagger(x_-)\,W_{hc}^\dagger(x)\,\xi(x)$
\cite{Hill:2002vw,Beneke:2002ni,Becher:2003qh} is the gauge-invariant (under 
both soft and hard-collinear gauge transformations) effective-theory field for 
a massless quark after a decoupling transformation has been applied, which 
removes the interactions of soft gluons with hard-collinear fields in the 
leading-order SCET Lagrangian \cite{Bauer:2001yt}. In the absence of such 
interactions the hard-collinear Lagrangian is equivalent to the conventional 
QCD Lagrangian, and we can rewrite the propagator in terms of standard QCD 
fields as 
\begin{equation}\label{eq:Jqcd}
   \frac{\sla{n}}{2}\,\nb\cdot p\,{\cal J}(p^2,\mu)
   = \int d^4x\,e^{-ip\cdot x}\,\langle 0\,|\,{\rm T}\left\{
   \frac{\sla{n}\sla{\nb}}{4}\,W^\dagger(0)\,\psi(0)\,\overline\psi(x)\,W(x)\,
   \frac{\sla{\nb}\sla{n}}{4} \right\} |\,0\rangle \,.
\end{equation} 
The quark fields are multiplied by Wilson lines 
\begin{equation}
   W(x) = {\rm\bf P}\,\exp\left( ig\int_{-\infty}^0\!ds\,
   \nb\cdot A(x+s\bar{n}) \right) ,
\end{equation}
which render the expression (\ref{eq:Jqcd}) gauge invariant. Note that the 
Wilson lines are absent in the light-cone gauge $\nb\cdot A=0$. For this 
reason the function ${\cal J}$ is sometimes referred to as the quark 
propagator in axial gauge. Lorentz invariance dictates that the QCD propagator 
in the presence of these Wilson lines contains two Dirac structures 
proportional to $\sla{p}$ and $\sla{\nb}$. The Dirac matrices appearing to the 
left and right of the field operators in (\ref{eq:Jqcd}) project out the terms 
proportional to $\sla{p}$. The jet function $J$ is the discontinuity of the 
propagator, i.e.\
\begin{equation}
   J(p^2,\mu) = \frac{1}{\pi}\,{\rm Im}\left[ i{\cal J}(p^2,\mu) \right]
   = \delta(p^2) + {\cal O}(\alpha_s) \,.
\end{equation}
Finally, we calculate the function $j$ from the contour integral
\begin{equation}
   j\Big( \ln\frac{Q^2}{\mu^2},\mu \Big)
   = \int_0^{Q^2}\!dp^2\,J(p^2,\mu)
   = - \frac{1}{2\pi} \ointctrclockwise\limits_{|\ps|=Q^2}\!d\ps\,
   {\cal J}(p^2,\mu) \,.
\end{equation}

\begin{figure}
\begin{center}
\includegraphics[width=0.9\textwidth]{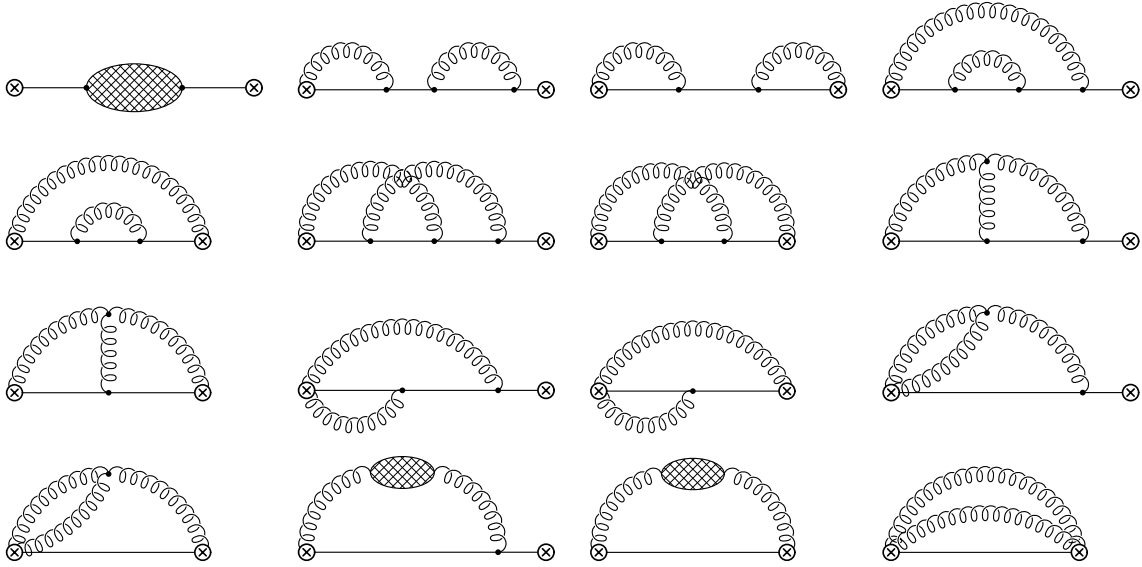}
\end{center}
\caption{\label{fig:jetgraphs}
Two-loop diagrams contributing to the jet function in QCD. Gluons emitted from 
the crossed circles originate from the Wilson lines. Not shown are additional 
diagrams resulting from mirror images in which the two external points are 
exchanged. The first diagram is the full fermion two-point function, not just 
the one-particle irreducible part.}
\end{figure}

Our calculation of the jet function employs the representation (\ref{eq:Jqcd}) 
of the function ${\cal J}(p^2,\mu)$ in terms of ordinary QCD quark and gluon 
fields. The relevant two-loop diagrams are shown in 
Figure~\ref{fig:jetgraphs}. Equally well, one could use the SCET Lagrangian 
together with  (\ref{eq:Jscet}) to perform the calculation. In this case 
diagrams in which a quark emits more than one gluon at the same vertex would 
also be present, in addition to the topologies shown in 
Figure~\ref{fig:jetgraphs}. Also, the analysis would be complicated by the 
fact that the SCET Feynman rules are more complicated that those of QCD. 

\subsection{Evaluation of the two-loop diagrams}
\label{sec:2loops}

We first discuss the evaluation of the bare quantity $j_{\rm bare}(Q^2)$ and 
later perform its renormalization. Let us begin by quoting the result for the 
one-loop master integral
\begin{equation}\label{1loop}
   \int\!d^dk\,\frac{(-1)^{-a-b-c}}{\left(k^2+i0\right)^a
                     \left[(k+p)^2+i0\right]^b \left(\nb\cdot k\right)^c}
   = i\pi^\frac{d}{2} \left( -p^2 - i0 \right)^{\frac{d}{2}-a-b} 
    \left( \nb\cdot p \right)^{-c} J(a,b,c) \,,
\end{equation}
with
\begin{equation}\label{eq:oneloopint}
   J(a,b,c)
   = \frac{\Gamma(\frac{d}{2}-b)\,\Gamma(\frac{d}{2}-a-c)\,
           \Gamma(a+b-\frac{d}{2})}%
          {\Gamma(a)\,\Gamma(b)\,\Gamma(d-a-b-c)} \,.
\end{equation}
At two-loop order, the most general integral we need is (omitting the 
``$+i0$'' terms for brevity)
\begin{eqnarray}\label{eq:twoloopint}
   \int\!d^dk\,\int\!d^dl\, &&\hspace{-0.5cm}
   \frac{(-1)^{-a_1-a_2-a_3-b_1-b_2-b_3-c_1-c_2}}%
        {\left(k^2\right)^{a_1} \left(l^2\right)^{a_2}
         \left[(k-l)^2\right]^{a_3} \left[(k+p)^2\right]^{b_1}
         \left[(l+p)^2\right]^{b_2} \left[(k+l+p)^2\right]^{b_3}
         \left(\nb\cdot k\right)^{c_1} \left(\nb\cdot l\right)^{c_2}}
    \nonumber\\
   &=& - \pi^d \left( -p^2 \right)^{d-a_1-a_2-a_3-b_1-b_2-b_3}
    \left( \nb\cdot p \right)^{-c_1-c_2} J(a_1,a_2,a_3,b_1,b_2,b_3,c_1,c_2)
    \,.
\end{eqnarray}
We use the same standard reduction techniques as in the two-loop calculation 
of the soft function \cite{Becher:2005pd} to express all integrals we need for 
the evaluation of the diagrams in Figure~\ref{fig:jetgraphs} in terms of four 
master integrals $M_n$. Introducing the dimensional regulator 
$\epsilon=2-d/2$, we obtain
\begin{eqnarray}\label{eq:masterints}
   M_1 &=& J(1,1,0,0,0,1,0,0)
    = \frac{\Gamma^3(1-\epsilon)\,\Gamma(2\epsilon-1)}{\Gamma(3-3\epsilon )} 
    \,, \nonumber\\
   M_2 &=& J(1,1,0,1,1,0,0,0) = J(1,1,0)^2
    = \left[
    \frac{\Gamma^2(1-\epsilon)\,\Gamma(\epsilon)}{\Gamma(2-2\epsilon )} 
    \right]^2 , \nonumber\\
   M_3 &=& J(1,0,1,1,1,0,0,1)
    = \frac{\Gamma(2\epsilon)}{\epsilon} \int_0^1\!dx \int_0^1\!dy\,
    \frac{(x\bar x y^2\bar y)^{-\epsilon}}{1-xy} \nonumber\\
   &&\hspace{3.545cm}
    = e^{-2\epsilon\gamma_E} \left( \frac{\pi^2}{12\epsilon^2}
    + \frac{7\zeta_3}{2\epsilon} + \frac{11\pi^4}{72} + {\cal O}(\epsilon) 
    \right) , \nonumber\\
   M_4 &=& J(1,1,0,1,1,1,1,1)
    = e^{-2\epsilon\gamma_E} \left( \frac{\pi^2}{3\epsilon^2}
    - \frac{7\zeta_3}{\epsilon} + \frac{23\pi^4}{360} + {\cal O}(\epsilon) 
    \right) ,
\end{eqnarray} 
where we use the shorthand notation $\bar x =1-x$ and $\bar y=1-y$. The 
evaluation of the first three integrals is straightforward. In the case of 
$M_3$ the double parameter integral resulting from the loop integrations can 
be expanded in $\epsilon$ without difficulty. The calculation of the master 
integral $M_4$, which is needed for the evaluation of the seventh graph in 
Figure~\ref{fig:jetgraphs}, is notably more complicated.

To tackle this last integral we use the Mellin-Barnes technique 
\cite{Smirnov:1999gc,Tausk:1999vh}. The basic strategy is to first introduce 
Feynman parameters to perform the loop integration and then introduce 
Mellin-Barnes parameters to carry out the Feynman parameter integrations. What 
makes this method powerful is that (after analytic continuation to 
$\epsilon\approx 0$) the Mellin-Barnes integrands can be Taylor expanded about 
$\epsilon=0$. To start, note that after performing the loop integral over $k$ 
using conventional Feynman parameters the result for $M_4$ can be written as
\begin{eqnarray}\label{eq:kint}
   M_4 &=& i\pi^{-\frac{d}{2}} \left( -p^2 \right)^{5-d}
    \left( \nb\cdot p \right)^2 \Gamma(3-{\textstyle\frac{d}{2}})
    \int\!d^dl\,\frac{1}{l^2\,(l+p)^2\,n\cdot l} \nonumber\\
   &\times& \int_0^1\!dx \int_0^1\!dy\,
    \frac{1}{\bar n\cdot p+\bar y\,\bar n\cdot l}\,
    \frac{1}{\left( -x^2 y\bar y l^2 - x\bar x\bar y (l+p)^2 - x\bar x y p^2
             \right)^{3-\frac{d}{2}}} \,.
\end{eqnarray}
We now introduce two Mellin-Barnes parameters via
\begin{equation}
   (A+B)^{-\alpha} = \frac{1}{2\pi i} \int_{c-i\infty}^{c+i\infty}\!dw\,
   A^w\,B^{-\alpha-w}\,\frac{\Gamma(-w)\,\Gamma(\alpha+w)}{\Gamma(\alpha)}
\end{equation}
to break apart the last denominator in (\ref{eq:kint}) into a product with 
factors $l^2$, $(l+p)^2$, and $p^2$. (We do not introduce a Mellin-Barnes 
parameter for the denominator $(\nb\cdot p+\bar y\,\nb\cdot l)$, as this would 
lead to ambiguities in the treatment of poles in the resulting light-cone 
propagators.) We then introduce additional Feynman parameters and perform the 
loop integration over $l$. We use a third Mellin-Barnes parameter to simplify 
the resulting Feynman parameter integrals, which can then be expressed in 
terms of $\Gamma$-functions. This leaves us with the following three-fold 
Mellin-Barnes integral:
\begin{eqnarray}\label{eq:MBrep}
   M_4 &=& \frac{1}{(2\pi i)^3}\,\prod_{i=1}^3
    \int_{c_i-i \infty}^{c_i+i \infty}\!\!dw_i\,
    \frac{\Gamma(-w_1)\,\Gamma(-w_2)\,\Gamma(-w_3)\,\Gamma(1+w_1)}%
         {\Gamma(-2\epsilon)\,\Gamma(1-2\epsilon+w_1+w_2+w_3)\,
          \Gamma(1-w_2)\,\Gamma(1-w_3)} \nonumber\\
   &\times& \Gamma(-\epsilon-w_2)\,\Gamma(\epsilon-w_2-w_3)\,
    \Gamma(1-\epsilon+w_1+w_3)\,\Gamma(1+\epsilon+w_2+w_3)\,
    \Gamma(1+w_1+w_2+w_3) \nonumber\\
   &\times& \Gamma(w_2-\epsilon)\,\Gamma(-1-\epsilon-w_1-w_3) \,.
\end{eqnarray}
In deriving this representation we have interchanged loop, Feynman, and 
Mellin-Barnes integrations. Careful inspection reveals that the representation 
(\ref{eq:MBrep}) is valid if the real parts of all $\Gamma$-functions are 
positive, i.e., if the poles of each $\Gamma$-function are either all to the 
left or all to the right of the integration contours. With the choice 
$c_1=-\frac14$, $c_2=-\frac18$, and $c_3=-\frac38$, this condition is 
fulfilled if $-\frac12<\epsilon<-\frac38$. Starting in the allowed range and 
increasing $\epsilon$, we see that at $\epsilon=-\frac38$ the first pole in 
$\Gamma(-1-\epsilon-w_1-w_3)$ crosses the contour from the right to the left. 
At $\epsilon=-\frac18$ the first pole in $\Gamma(w_2-\epsilon)$ crosses from 
the left to the right. The arguments of all other $\Gamma$-functions remain 
positive up to $\epsilon<\frac18$. To obtain a representation that is valid 
around $\epsilon=0$, one has to either deform the contours such that the 
crossings are avoided, or separately take into account the contributions of 
the poles that end up on the wrong side of the contours as $\epsilon\to 0$, as 
proposed in \cite{Tausk:1999vh}. The residues of these poles have again the 
form of Mellin-Barnes integrals, however with one integration less than the 
original expression (\ref{eq:MBrep}). One analyzes these contributions with 
the same method as the original integral and continues until one ends up with 
a representation which is valid around $\epsilon=0$. Once this is achieved, 
the integrands are expanded in $\epsilon$, the integration contours are 
closed, and the integrations are rewritten as sums over the residues of the 
poles. In fact, since the original integral includes a factor 
$1/\Gamma(-2\epsilon)=\order(\epsilon)$, only contributions which arise from 
poles that cross a contour need to be included in the limit $\epsilon\to 0$. 
Very recently, the continuation in $\epsilon$ with subsequent numerical 
evaluation of the integrals has been automatized 
\cite{Anastasiou:2005cb,Czakon:2005rk}. We have used the public code of 
\cite{Czakon:2005rk} to check our analytical result for $M_4$. 

As a further independent check of our result we have numerically evaluated the 
multi-dimen\-sional integral over Feynman parameters, which is obtained after 
performing the loop integration over $l$ in (\ref{eq:kint}) using conventional 
methods. As is evident from (\ref{eq:masterints}) the integral $M_4$ is 
divergent, and the divergences need to be isolated in order to perform the 
numerical evaluation. We use the method of sector decomposition 
\cite{Hepp:1966eg,Binoth:2000ps,Anastasiou:2003gr}, which allows one to 
systematically disentangle overlapping singularities in Feynman integrals. 
This procedure splits the integral into a large number of terms in which all 
singularities are factorized. Because it leads to large algebraic expressions, 
the sector decomposition is performed using computer algebra. After numerical 
integration of the resulting expressions we reproduce the analytical result 
for the integral $M_4$ with a numerical precision of better than 1 part in 
$10^6$.

With the integrals at hand, the evaluation of the two-loop diagrams in 
Figure~\ref{fig:jetgraphs} is straightforward. We write each diagram as a sum 
of integrals of the form (\ref{eq:twoloopint}) and express those in terms of 
the four master integrals $M_1,\dots,M_4$. Summing up the results for the 
individual diagrams, we obtain
\begin{eqnarray}\label{sbare}
   j_{\rm bare}(Q^2)
   &=& 1 + \frac{Z_\alpha\alpha_s}{4\pi}
    \left( \frac{Q^2}{\mu^2} \right)^{-\epsilon} C_F \left[
    \frac{4}{\epsilon^2} + \frac{3}{\epsilon} + 7 - \pi^2
    + \left( 14 - \frac{3\pi^2}{4} - \frac{28}{3}\,\zeta_3 \right) \epsilon
    \right. \nonumber\\
   &&\left.\hspace{3.6cm}\mbox{}+ \left( 28 - \frac{7\pi^2}{4}
    - \frac{\pi^4}{24} - 7\zeta_3 \right) \epsilon^2 + {\cal O}(\epsilon^3)
    \right] \nonumber\\
   &&\mbox{}+ \left( \frac{Z_\alpha\alpha_s}{4\pi} \right)^2
    \left( \frac{Q^2}{\mu^2} \right)^{-2\epsilon} C_F\,
    \Big[ C_F K_F(\epsilon) + C_A K_A(\epsilon) + T_F n_f K_f(\epsilon) \Big]
    + \dots \,,
\end{eqnarray}
where
\begin{equation}
   Z_\alpha = 1 - \beta_0\,\frac{\alpha_s}{4\pi\epsilon} + \dots
   = 1 - \left( \frac{11}{3}\,C_A - \frac43\,T_F n_f \right)\,
   \frac{\alpha_s}{4\pi\epsilon}  + \dots \,.
\end{equation}
Here $\alpha_s\equiv\alpha_s(\mu)$ is the renormalized coupling. Note that the 
bare jet function is scale independent, since the bare coupling 
$Z_\alpha\alpha_s\mu^{2\epsilon}$ does not depend on the renormalization 
scale. The two-loop coefficients are
\begin{eqnarray}\label{eq:jbare}
   K_F(\epsilon)
   &=& \frac{8}{\epsilon^4} + \frac{12}{\epsilon^3}
    + \left( \frac{65}{2} - \frac{16\pi^2}{3} \right) \frac{1}{\epsilon^2}
    + \left( \frac{311}{4} - 9\pi^2 - \frac{124}{3}\,\zeta_3 \right)
    \frac{1}{\epsilon} \nonumber\\
    &&\mbox{}+ \frac{1437}{8} - \frac{301\pi^2}{12} + \frac{113\pi^4}{90}
     - 86\zeta_3 + {\cal O}(\epsilon) \,, \nonumber\\
   K_A(\epsilon)
   &=& \frac{11}{3\epsilon^3}
    + \left( \frac{233}{18} - \frac{\pi^2}{3} \right) \frac{1}{\epsilon^2}
    + \left( \frac{4541}{108} - \frac{55\pi^2}{18} - 20\zeta_3 \right)
    \frac{1}{\epsilon} \nonumber\\
   &&\mbox{}+ \frac{86393}{648} - \frac{1129\pi^2}{108} - \frac{17\pi^4}{180} 
    - \frac{514}{9}\,\zeta_3 + {\cal O}(\epsilon) \,, \nonumber\\
   K_f(\epsilon)
   &=& -\frac{4}{3\epsilon^3} - \frac{38}{9\epsilon^2}
    + \left( -\frac{373}{27} + \frac{10\pi^2}{9} \right) \frac{1}{\epsilon}
    - \frac{7081}{162} + \frac{95\pi^2}{27} + \frac{128}{9}\,\zeta_3
    + {\cal O}(\epsilon) \,.
\end{eqnarray}
We have checked that the divergences of the first diagram in 
Figure~\ref{fig:jetgraphs} (together with the appropriate one-loop counter 
term, and accounting for the renormalization of the gauge parameter) reproduce 
the known result for the two-loop quark wave-function renormalization 
\cite{Egorian:1978zx}. A stringent check of the remaining diagrams with gluon 
emissions from the Wilson lines is that their divergences must cancel against 
the jet-function renormalization factor, which we will now derive.

\subsection{Renormalization of the jet function}

The renormalization of the bare jet function proceeds in complete analogy to 
that of the bare soft function discussed in \cite{Becher:2005pd}, to which we 
refer the reader for a more detailed discussion. The procedure is more 
complicated than in conventional applications of renormalization owing to the 
presence of Sudakov double logarithms.

We define an operator renormalization factor for the jet function via
\begin{equation}\label{Jrge}
   J(\ps,\mu) = \int d\psp\,Z(\ps,\psp,\mu)\,J_{\rm bare}(\psp) \,,
\end{equation}
where $Z$ absorbs the UV divergences of the bare jet function, such that the
renormalized jet function is finite in the limit $\epsilon\to 0$. In the 
$\overline{\rm MS}$ regularization scheme, we have
\begin{equation}
   Z(\ps,\psp,\mu) = \delta(\ps-\psp)
   + \sum_{k=1}^\infty\,\frac{1}{\epsilon^k}\,Z^{(k)}(\ps,\psp,\mu) \,.
\end{equation}
The relations
\begin{equation}\label{rela}
   \gamma_{\rm jet}
   = 2\alpha_s\,\frac{\partial Z^{(1)}}{\partial\alpha_s} \,, \qquad 
   2\alpha_s\,\frac{\partial Z^{(n+1)}}{\partial\alpha_s}
   = 2\alpha_s\,\frac{\partial Z^{(1)}}{\partial\alpha_s}\otimes Z^{(n)}
    + \beta(\alpha_s)\,\frac{\partial Z^{(n)}}{\partial\alpha_s}
    + \frac{\partial Z^{(n)}}{\partial\ln\mu} \,,
\end{equation}
with $n\geq 1$, connect the coefficient of the $1/\epsilon$ pole in the $Z$
factor to the anomalous dimension and moreover imply a set of consistency 
conditions among the coefficients of the higher pole terms. The symbol 
$\otimes$ represents a convolution in $p^2$. The term 
$\partial Z^{(n)}/\partial\ln\mu$ arises because the $Z$ factor depends both 
implicitly (via the renormalized coupling constant) and explicitly (via 
Sudakov logarithms contained in star distributions) on the renormalization 
scale \cite{Becher:2005pd}. 

To all orders in perturbation theory the anomalous-dimension kernel of the 
jet function has the form 
\begin{equation}\label{gammaJ}
   \gamma_{\rm jet}(\ps,\psp,\mu)
   = 2\Gamma_{\rm cusp}(\alpha_s) \left( \frac{1}{\ps-\psp}
   \right)_{\!*}^{\![\mu^2]} + 2\gamma^J(\alpha_s)\,\delta(\ps-\psp) \,,
\end{equation}
where $\Gamma_{\rm cusp}$ is the cusp anomalous dimension associated with the 
Sudakov double logarithms, while $\gamma^J$ controls the single-logarithmic
evolution of the jet function. The definition of the star distribution can be 
found in \cite{Bosch:2004th}. The corresponding integro-differential evolution 
equation reads
\begin{equation}\label{Jreg}
   \frac{dJ(p^2,\mu)}{d\ln\mu}
   = - \left[ 2\Gamma_{\rm cusp}\,\ln\frac{p^2}{\mu^2} + 2\gamma^J \right]
   J(p^2,\mu) - 2\Gamma_{\rm cusp} \int_0^{p^2}\!dp^{\prime 2}\,
   \frac{J(\psp,\mu)-J(p^2,\mu)}{p^2-\psp} \,.
\end{equation} 
We have derived relation (\ref{gammaJ}) by requiring that the 
$\bar B\to X_s\gamma$ decay rate be renormalization-group invariant and using 
the known evolution equations for the soft function \cite{Grozin:1994ni} and 
for the hard matching coefficient \cite{Bauer:2000yr,Becher:2003kh}. Denoting 
by $Z_{[n]}$ the coefficient of $(\alpha_s/4\pi)^n$ in $Z(\ps,\psp,\mu)$, we 
obtain from (\ref{rela})
\begin{eqnarray}
   Z_{[0]} &=& \delta(\ps-\psp) \,, \nonumber\\
   Z_{[1]} &=& \delta(\ps-\psp)
    \left( -\frac{\Gamma_0}{\epsilon^2} + \frac{\gamma^J_0}{\epsilon} \right)
    + \frac{\Gamma_0}{\epsilon}
    \left( \frac{1}{\ps-\psp} \right)_{\!*}^{\![\mu^2]} , \nonumber\\
   Z_{[2]} &=& \delta(\ps-\psp)
    \left[ \frac{\Gamma_0^2}{2\epsilon^4}
    - \frac{\Gamma_0(\gamma^J_0 - \frac34\beta_0)}{\epsilon^3}
    + \left( \frac{\gamma^J_0(\gamma^J_0-\beta_0)}{2} - \frac{\Gamma_1}{4}
    - \frac{\pi^2}{12}\,\Gamma_0^2 \right) \frac{1}{\epsilon^2}
    + \frac{\gamma^J_1}{2\epsilon} \right] \nonumber\\
   &&\mbox{}+ \left[ - \frac{\Gamma_0^2}{\epsilon^3}
    + \frac{\Gamma_0(\gamma^J_0 - \frac12\beta_0)}{\epsilon^2}
    + \frac{\Gamma_1}{2\epsilon} \right]
    \left( \frac{1}{\ps-\psp} \right)_{\!*}^{\![\mu^2]}
    + \frac{\Gamma_0^2}{\epsilon^2}
    \left( \frac{\ln\frac{\ps-\psp}{\mu^2}}{\ps-\psp}
    \right)_{\!*}^{\![\mu^2]} ,
\end{eqnarray}
where the expansion coefficients of the anomalous dimensions and 
$\beta$-function are defined as
\begin{eqnarray}
   \Gamma_{\rm cusp}(\alpha_s)
   &=& \sum_{n=0}^\infty\,\Gamma_n \left( \frac{\alpha_s}{4\pi} \right)^{n+1}
    , \qquad
   \gamma^J(\alpha_s)
   = \sum_{n=0}^\infty\,\gamma_n \left( \frac{\alpha_s}{4\pi} \right)^{n+1}
    , \nonumber\\
   \beta(\alpha_s)
   &=& \frac{d\alpha_s}{d\ln\mu} = -2\alpha_s
    \sum_{n=0}^\infty\,\beta_n \left( \frac{\alpha_s}{4\pi} \right)^{n+1} .
\end{eqnarray}
The expression for the two-loop cusp anomalous dimension can be found, e.g., 
in our previous paper \cite{Becher:2005pd}. The two-loop anomalous dimension 
of the jet function has never been calculated directly, but it was inferred in 
\cite{Neubert:2004dd} from existing two-loop results for jet-function moments 
in deep-inelastic scattering \cite{Vogt:2000ci}. In that way one obtains
\begin{eqnarray}\label{g1J}
   \gamma_0^J &=& -3 C_F \,, \\
   \gamma_1^J &=& C_F^2 \left( - \frac{3}{2} + 2\pi^2 - 24\zeta_3 \right) 
    + C_F C_A \left( - \frac{1769}{54} - \frac{11\pi^2}{9} + 40\zeta_3 \right)
    + C_F T_F n_f \left( \frac{242}{27} + \frac{4\pi^2}{9} \right) \,.
    \nonumber
\end{eqnarray}

It is straightforward to show that the function $j(\ln\frac{Q^2}{\mu^2},\mu)$ 
obeys the same evolution equation as the original jet function $J(p^2,\mu)$, 
i.e.\ 
\begin{equation}
   j\Big( \ln\frac{Q^2}{\mu^2},\mu \Big)
   = \int_0^{Q^2}\!dQ^{\prime 2}\,Z(Q^2,Q^{\prime 2},\mu)\,
   j_{\rm bare}(Q^{\prime 2}) \,,
\end{equation}
where $j_{\rm bare}(Q^2)$ is the quantity we have calculated in 
Section~\ref{sec:2loops}. Expanding this relation in perturbation theory we 
obtain $j_{[0]}=j_{[0]}^{\rm bare}=1$ and 
\begin{equation}
   j_{[1]} = j_{[1]}^{\rm bare} + Z_{[1]}\otimes j_{[0]}^{\rm bare} \,,
    \qquad
   j_{[2]} = j_{[2]}^{\rm bare} + Z_{[1]}\otimes j_{[1]}^{\rm bare}
    + Z_{[2]}\otimes j_{[0]}^{\rm bare} \,.
\end{equation}
The first term on the right-hand side in each equation corresponds to the 
contribution (\ref{sbare}) obtained from the loop diagrams. The remaining 
terms correspond to the counter-term contributions. Explicitly, we find
\begin{eqnarray}\label{opsCT}
   j_{[1]}^{\rm C.T.}
   &=& -\frac{\Gamma_0}{\epsilon^2} + \frac{\gamma^J_0}{\epsilon}
    + \frac{\Gamma_0}{\epsilon}\,\ln\frac{Q^2}{\mu^2} \,, \nonumber\\
   j_{[2]}^{\rm C.T.}
   &=& \left[ -\frac{\Gamma_0}{\epsilon^2}
    + \frac{\gamma^J_0}{\epsilon} + \frac{\Gamma_0}{\epsilon} \left(
    \ln\frac{Q^2}{\mu^2} - \gamma_E - \psi(1-\epsilon) \right) \right]
    j_{[1]}^{\rm bare}(Q^2) \nonumber\\
   &&\mbox{}+ \frac{\Gamma_0^2}{2\epsilon^4}
    - \frac{\Gamma_0(\gamma^J_0 - \frac34\beta_0)}{\epsilon^3}
    + \left( \frac{\gamma^J_0(\gamma^J_0-\beta_0)}{2} - \frac{\Gamma_1}{4}
    - \frac{\pi^2}{12}\,\Gamma_0^2 \right) \frac{1}{\epsilon^2}
    + \frac{\gamma^J_1}{2\epsilon} \nonumber\\
   &&\mbox{}+ \left[ - \frac{\Gamma_0^2}{\epsilon^3}
    + \frac{\Gamma_0(\gamma^J_0 - \frac12\beta_0)}{\epsilon^2}
    + \frac{\Gamma_1}{2\epsilon} \right] \ln\frac{Q^2}{\mu^2}
    + \frac{\Gamma_0^2}{2\epsilon^2}\,\ln^2\frac{Q^2}{\mu^2} \,.
\end{eqnarray}
Together with the results for the bare one-loop jet function from 
(\ref{sbare}) this yields explicit expressions for the counter terms. When 
adding these contributions to the bare jet function we find that all 
$1/\epsilon^n$ pole terms cancel, so that the limit $\epsilon\to 0$ can now be 
taken. This implies, in particular, that we confirm by direct calculation the 
expression for the anomalous-dimension coefficient $\gamma_1^J$ given 
in~(\ref{g1J}).

\subsection{Results}

The logarithmic terms in the renormalized jet function have been determined in 
\cite{Neubert:2005nt} by solving the renormalization-group equation 
perturbatively. At two-loop order, it was found that
\begin{eqnarray}\label{jrg}
   j(L,\mu) &=& 1 + \frac{\alpha_s(\mu)}{4\pi} \left[
    b_0^{(1)} + \gamma_0^J L + \frac{\Gamma_0}{2}\,L^2 \right] \nonumber\\
   &&\mbox{}+ \left( \frac{\alpha_s(\mu)}{4\pi} \right)^2 \Bigg[
    b_0^{(2)} + \left( b_0^{(1)} (\gamma_0^J-\beta_0) + \gamma_1^J
    - \frac{\pi^2}{6}\,\Gamma_0\gamma_0^J + \zeta_3\,\Gamma_0^2 \right) L
    \nonumber\\
   &&\quad\mbox{}+ \frac12 \left( \gamma_0^J (\gamma_0^J-\beta_0)
    + b_0^{(1)}\,\Gamma_0 + \Gamma_1
    - \frac{\pi^2}{6}\,\Gamma_0^2 \right) L^2
    + \frac{\Gamma_0}{2} \left( \gamma_0^J - \frac{\beta_0}{3} \right) L^3
    + \frac{\Gamma_0^2}{8}\,L^4 \Bigg] \,. \qquad
\end{eqnarray}
Our results for the logarithmic terms agree with the above expression. The 
one-loop coefficient $b_0^{(1)}$ was derived in 
\cite{Bauer:2003pi,Bosch:2004th}. The main new result is $b_0^{(2)}$, the 
constant term at two-loop order. We obtain
\begin{eqnarray}\label{b0n}
   b_0^{(1)} &=& (7-\pi^2)\,C_F \,, \nonumber\\
   b_0^{(2)} &=& C_F^2 \left( \frac{205}{8} - \frac{67\pi^2}{6}
    + \frac{14\pi^4}{15} - 18\zeta_3 \right) 
    + C_F C_A \left( \frac{53129}{648} - \frac{208\pi^2}{27} 
    - \frac{17\pi^4}{180} - \frac{206}{9}\,\zeta_3 \right) \nonumber\\
   &&\mbox{}+ C_F T_F n_f \left( - \frac{4057}{162} + \frac{68\pi^2}{27}
    + \frac{16}{9}\,\zeta_3 \right) .
\end{eqnarray}
It is interesting to compare the exact answer for the coefficient $b_0^{(2)}$ 
with the approximation obtained by keeping only the terms of order 
$\beta_0\alpha_s^2$. In the absence of exact two-loop results it is sometimes 
argued that the $\beta_0\alpha_s^2$ terms constitute the dominant part of the 
complete two-loop correction. In the present case, we obtain for $N_c=3$ 
colors $b_0^{(2)}\approx-16.25 x-128.78\approx -145.04$, where 
$x=\frac{3}{25}\,\beta_0=1$ for $n_f=4$ light flavors. Keeping only the 
$\beta_0\alpha_s^2$ terms would give $-16.25$, which is off by an order of 
magnitude. This illustrates the importance of performing exact two-loop 
calculations.

\begin{figure}
\begin{center}
\includegraphics[width=0.9\textwidth]{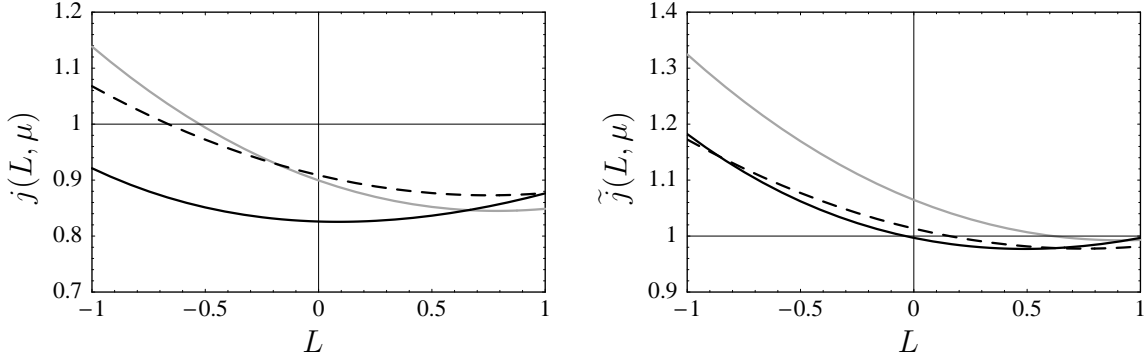}
\end{center}
\vspace{-0.5cm}
\caption{ \label{fig:jetfunction}
One- and two-loop predictions for the jet functions $j(L,\mu)$ and 
$\widetilde j(L,\mu)$ evaluated at $\alpha_s(\mu)=0.3$. The dashed lines show 
the one-loop results, while the solid lines give the complete two-loop results 
derived in the present work. The gray lines are obtained if only the 
$\beta_0\alpha_s^2$ terms are kept in the two-loop contributions.}
\end{figure}

We now briefly discuss the impact of our results for phenomenology. Besides 
the jet function $j(L,\mu)$ itself, it is useful to consider a related 
function $\widetilde j(L,\mu)$ obtained by replacing the $n$-th power of $L$ 
in (\ref{sexp}) with an $n$-th order polynomial, $L^n\to I_n(L)$, where at 
two-loop order we need
\begin{eqnarray}
    I_1(x) &=& x \,, \hspace{2.0cm}
     I_3(x) = x^3 + \frac{\pi^2}{2}\,x - 2\zeta_3 \,, \nonumber\\
    I_2(x) &=& x^2 + \frac{\pi^2}{6} \,, \qquad
     I_4(x) = x^4 + \pi^2 x^2 - 8\zeta_3\,x + \frac{3\pi^4}{20} \,.
\end{eqnarray}
The function $\widetilde j$ enters in the factorization formula for the 
partial $\bar B\to X_s\gamma$ decay rate with a cut on photon energy 
\cite{Neubert:2005nt}. For the case of $N_c=3$ colors and $n_f=4$ light quark 
flavors we get
\begin{eqnarray}
   j(L,\mu)
   &\approx& 1 + \left( - 0.304 - 0.318 L + 0.212 L^2 \right) \alpha_s(\mu)
    \nonumber\\
   &&\mbox{}+ \left( - 0.918 + 0.926 L + 0.079 L^2 - 0.114 L^3 + 0.023 L^4 
    \right) \alpha_s^2(\mu) + \dots \,, \nonumber\\
   \widetilde j(L,\mu)
   &\approx& 1 + \left( 0.045 - 0.318 L + 0.212 L^2 \right) \alpha_s(\mu)
    \nonumber\\
   &&\mbox{}+ \left( - 0.185 + 0.145 L + 0.301 L^2 - 0.114 L^3 + 0.023 L^4 
    \right) \alpha_s^2(\mu) + \dots \,.
\end{eqnarray}
The two-loop corrections to $j$ are very large and, for realistic parameter 
values, can even dominate over the one-loop corrections. However, the two-loop 
corrections are much smaller for the function $\widetilde j$. 
Figure~\ref{fig:jetfunction} shows the dependence of the two jet functions on 
$L=\ln(Q^2/\mu^2)$ for a fixed scale $\mu\approx 2$\,GeV chosen such that 
$\alpha_s(\mu)=0.3$, corresponding to a renormalization point appropriate for 
the calculation of the partial $\bar B\to X_s\gamma$ decay rate with a cut 
$E_\gamma>1.8$\,GeV. The two-loop effects calculated in this work impact the 
jet function $j$ at the 10\% level, while their effect on the function 
$\widetilde j$ is at the level of 2\% or less. The latter finding suggests a 
good convergence of the perturbative expansion at the intermediate scale in 
the analysis of $\bar B\to X_s\gamma$ decay. In addition to the one- and 
two-loop predictions, the figure also displays the results obtained if only 
terms of order $\beta_0\alpha_s^2$ are kept in the two-loop coefficients. In 
both cases this provides a poor approximation to the exact two-loop results. 
We also note that the jet functions by themselves are not 
renormalization-group invariant, so it is meaningless to study their 
dependence on the scale $\mu$ for fixed $Q^2$. In physical results such as the 
expression for the $\bar B\to X_s\gamma$ decay rate and photon-energy moments 
given in \cite{Neubert:2004dd,Neubert:2005nt}, the scale dependence of the jet 
function cancels against that of other renormalization-group functions. 

\section{Moments of the jet function}
\label{sec:DIS}

In the analysis of hard QCD processes such as deep-inelastic scattering it is 
often convenient to introduce moments of the jet function defined as (see 
e.g.\ \cite{Catani:1990rr})
\begin{equation}\label{JNdef}
   J_N(Q^2,\mu) = \int_0^{Q^2}\!dp^2 \left( 1 - \frac{p^2}{Q^2} \right)^{N-1}
   J(p^2,\mu) \,.
\end{equation} 
Whereas in inclusive $B$ decays the scale $Q^2$ setting the upper integration 
limit in (\ref{jdefsdef}) is an intermediate (hard-collinear) scale, 
$Q^2\ll m_b^2$, which is of order the invariant mass squared of the 
final-state hadronic jet, the variable $Q^2$ in (\ref{JNdef}) is set by a 
characteristic hard scale of the process. In the large-$N$ limit the integral 
receives leading contributions only from the region $p^2\sim Q^2/N\ll Q^2$. 
The scale $Q^2/N$ is the analog of the intermediate scale in 
$\bar B\to X_s\gamma$ decay.

Using an integration by parts, it is straightforward to express the 
jet-function moments in terms of integrals over the function $j$ calculated at 
two-loop order in the present work. We obtain
\begin{equation}
   J_1(Q^2,\mu) = j\Big( \ln\frac{Q^2}{\mu^2},\mu\Big) \,, \qquad
   J_N(Q^2,\mu) = (N-1) \int_0^1\!dx\,(1-x)^{N-2}\,
    j\Big( \ln\frac{xQ^2}{\mu^2},\mu\Big) \,,
\end{equation}
where the second relation holds for $N\ge 2$. It follows from (\ref{Jreg}) 
that the moments obey the evolution equation
\begin{eqnarray}\label{complicated}
   \frac{dJ_N(Q^2,\mu)}{d\ln\mu}
   &=& - \left[ 2\Gamma_{\rm cusp} \left( \ln\frac{Q^2}{\mu^2} - H_{N-1}
    \right) + 2\gamma^J \right] J_N(Q^2,\mu) \nonumber\\
   &&\mbox{}- 2\Gamma_{\rm cusp} \int_0^{Q^2}\!dp^2
    \left( 1 - \frac{p^2}{Q^2} \right)^{N-1}
    \ln\left( 1 - \frac{p^2}{Q^2} \right) J(p^2,\mu) \,,
\end{eqnarray}
where $H_{N-1}=\sum_{n=1}^{N-1} \frac{1}{n}$ is the harmonic number. These 
results simplify greatly in the large-$N$ limit. We find that the moments are 
given by 
\begin{equation}
   J_N(Q^2,\mu)
   = \widetilde j\Big( \ln\frac{Q^2}{e^{\gamma_E} N\mu^2},\mu \Big)
   + \order\left( \frac{1}{N} \right) ,
\end{equation}
and that they obey the {\em local\/} evolution equation
\begin{equation}\label{JNrge}
   \frac{dJ_N(Q^2,\mu)}{d\ln\mu}
   = - \left[ 2\Gamma_{\rm cusp}\,\ln\frac{Q^2}{e^{\gamma_E} N\mu^2}
   + 2\gamma^J \right] J_N(Q^2,\mu) + \order\left( \frac{1}{N} \right) .
\end{equation}
In deriving this result we have used that the second line in 
(\ref{complicated}) is suppressed in the large-$N$ limit, since 
$p^2/Q^2=\order(1/N)$ in the argument of the logarithm. This local evolution 
equation can be integrated using standard techniques.

\section{Conclusions}

We have calculated the two-loop expression for the jet function $j(L,\mu)$ 
defined in terms of an integral over the hard-collinear quark propagator in 
soft-collinear effective theory. This quantity is a necessary ingredient for 
the NNLO evaluation of the $\bar B\to X_s\gamma$ decay rate with a cut on the 
photon energy. Moreover, since the jet function is universal, it appears in 
many other applications of perturbative QCD. The results obtained in the 
present work, when combined with \cite{Becher:2005pd}, provide a complete 
description of low-scale effects in the analysis of the partial 
$\bar B\to X_s\gamma$ decay rate at NNLO in renormalization-group improved 
perturbation theory. A detailed study of the phenomenological impact of these 
effects will be presented elsewhere.

\subsection*{Acknowledgments}

We thank Frank Petriello for discussions and for numerically evaluating the 
master integral $M_4$. The research of T.B.\ was supported by the Department 
of Energy under Grant DE-AC02-76CH03000. The research of M.N.\ was supported 
by the National Science Foundation under Grant PHY-0355005. Fermilab is 
operated by Universities Research Association Inc., under contract with the 
U.S.\ Department of Energy.

\end{document}